\documentclass[prb,twocolumn,superscriptaddress]{revtex4}
\usepackage{graphicx}
\usepackage{amssymb}
\usepackage{amsmath}
\usepackage{graphicx}
\usepackage{epsfig}
\usepackage{array}
\usepackage{color}
\usepackage{ulem}
\usepackage{multirow}
\makeatletter
\renewcommand{\section}{\@startsection{section}{1}{0mm} 
  {-\baselineskip}{0.5\baselineskip}{\bf\leftline}} 
\makeatother 

\begin{document}

\title{Mechanism of the insulator-to-metal transition and superconductivity\\ in the spin liquid candidate NaYbSe$_2$ under pressure}

\author{Yuanji Xu}
\affiliation{Beijing National Laboratory for Condensed Matter Physics and
Institute of Physics, Chinese Academy of Sciences, Beijing 100190, China}
\author{Yutao Sheng}
\affiliation{Beijing National Laboratory for Condensed Matter Physics and
Institute of Physics, Chinese Academy of Sciences, Beijing 100190, China}
\affiliation{University of Chinese Academy of Sciences, Beijing 100049, China}
\author{Yi-feng Yang}
\email[]{yifeng@iphy.ac.cn}
\affiliation{Beijing National Laboratory for Condensed Matter Physics and
Institute of Physics, Chinese Academy of Sciences, Beijing 100190, China}
\affiliation{University of Chinese Academy of Sciences, Beijing 100049, China}
\affiliation{Songshan Lake Materials Laboratory, Dongguan, Guangdong 523808, China}

\date{\today}

\begin{abstract}
The quantum spin liquid candidate NaYbSe$_2$ was recently reported to exhibit a Mott transition under pressure. Superconductivity was observed in the high-pressure metallic phase, raising the question concerning its relation with the low-pressure quantum spin liquid ground state. Here we combine the density functional theory and the dynamical mean-field theory to explore the underlying mechanism of the insulator-to-metal transition and superconductivity and establish an overall picture of its electronic phases under pressure. Our results suggest that NaYbSe$_2$ is a charge-transfer insulator at ambient pressure. Upon increasing pressure, however, the system first enters a semi-metallic state with incoherent Kondo scattering against coexisting localized Yb-$4f$ moments, and then turns into a heavy fermion metal. In between, there may exist a delocalization quantum critical point responsible for the observed non-Fermi liquid region with linear-in-$T$ resistivity. The insulator-to-metal transition is therefore a two-stage process. Superconductivity emerges in the heavy fermion phase with well-nested Yb-4$f$ Fermi surfaces, suggesting that spin fluctuations may play a role in the Cooper pairing. NaYbSe$_2$ might therefore be the 3rd Yb-based heavy-fermion superconductor with a very ``high" $T_c$ than most heavy fermion superconductors.
\end{abstract}

\maketitle

\section*{INTRODUCTION}
Quantum spin liquid (QSL) is a highly entangled state with nontrivial topological excitations \cite{Zhou2017,Savary2017,Balents2010,Broholm2020}. It has attracted intensive interest in modern condensed matter physics and is believed to describe the ground state of the Mott insulator and, upon doping, can induce high-temperature superconductivity as in cuprates \cite{Anderson1973,Anderson1987,Wen1991,Kitaev2006,Wen2019}. Tremendous efforts have thus been devoted to exploring candidate spin liquid compounds \cite{Kelly2016,Kurosaki2005,Shimizu2016,Powell2011,Clancy2018,Podolsky2009} and searching for potential sign of superconductivity via chemical substitution \cite{Wen2006,Lee2008}. However, doping typically introduces disorder that may obscure the investigation of such delicate intrinsic properties.

Recently, superconductivity has been reported by pressurizing the QSL candidate NaYbSe$_{2}$ \cite{Jia2020,Zhang2020}. In contrast to YbMgGaO$_{4}$ which has the same structure (space group $R$-$3mH$) but with Mg/Ga disorders \cite{Li2015,Zhu2017,Yuesheng2015,Li2016,Shen2016,Li2017,Zhu2018,Kimchi2018}, NaYbSe$_2$ contains no intrinsic site disorder and has arguably the simplest crystal structure and chemical formula among all existing QSL candidates \cite{Gray2003}. At ambient pressure, analyses of transport and neutron data found an insulating charge gap of 1.9 eV \cite{Liu2018} and a crystalline excitation energy of about 15.7 meV \cite{Ranjith2019,Dai2021,Zhang2021}. No magnetic or structural transition was observed down to 50 mK \cite{Liu2018,Ranjith2019,Dai2021}. Neutron scattering measurement has revealed the signature of magnetic excitation continuum and supported a QSL ground state \cite{Dai2021}. The low-energy physics may thus be described by an effective spin-$1 \over 2$ model on a perfect triangular lattice formed by Yb ions. At 11 GPa, a structural transition was reported \cite{Jia2020}. The high-pressure structure has a lower symmetry (space group $P$-$3m1$) with two inequivalent Yb ions of different Yb-Se distances \cite{Jia2020}. Upon further increasing pressure, an insulator-to-metal transition appears at about 58.9 GPa and superconductivity emerges over a wide pressure range above 103.4 GPa with a maximum $T_c$ of about 8 K \cite{Jia2020}. The insulator-to-metal transition was claimed to be of Mott type, raising the question concerning the relationship between the low-pressure QSL and the high-pressure superconductivity.

\begin{figure}[t]
\begin{center}
\includegraphics[width=0.48\textwidth]{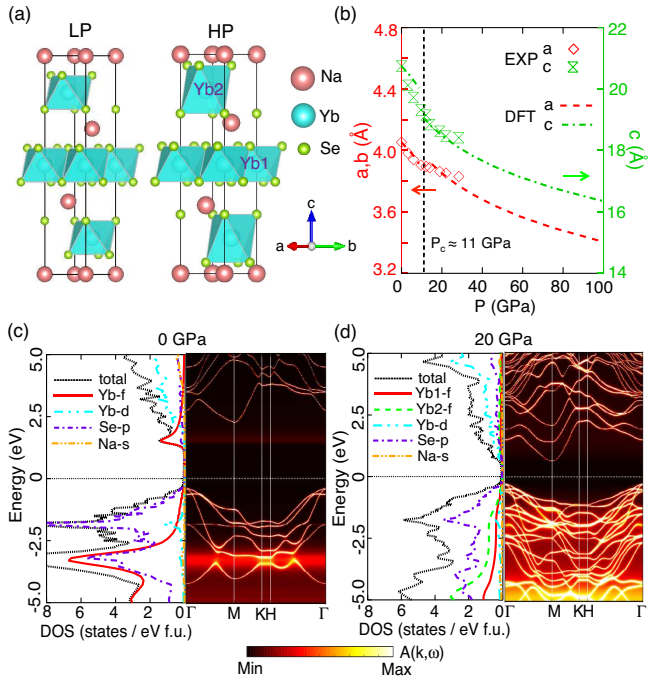}
\caption{{\bf Lattice structure and low-pressure electronic structures of NaYbSe$_2$.} {\bf a} Comparison of the crystal structures at low (LP) and high pressures (HP). {\bf b} Optimized lattice parameters as a function of pressure compared with experiment \cite{Jia2020,Zhang2020}. {\bf c, d} Orbital-resolved density of states and the spectral function at ambient pressure and 20 GPa obtained by DFT+DMFT at 10 K.}
\label{fig1}
\end{center}
\end{figure}

To clarify the nature of the ``Mott" transition and the superconducting pairing mechanism, we investigate here the electronic structure of NaYbSe$_{2}$ and its pressure evolution using the density functional theory (DFT) plus the dynamical mean-field theory (DMFT) calculations \cite{Kotliar2006,Georges1996,Haule2010,Held2008}. We find that NaYbSe$_2$ is a charge-transfer insulator at ambient pressure, with the charge gap formed between Se-$3p$ bands and Yb-$4f$ upper-Hubbard bands. The charge gap is closed near about 60 GPa, giving rise to an intermediate semi-metallic phase with small electron and hole pockets from Yb-$5d$ and Se-$3p$ bands, respectively. The Yb-$4f$ electrons remain localized at this pressure and the system is a typical low carrier density Kondo system, which explains the logarithmic temperature dependence of the resistivity observed in transport measurements. Above 73 GPa, the Yb-$4f$ moments are fully screened on one (Yb1) of the two inequivalent Yb-ions, giving rise to flat heavy electron bands near the Fermi energy with sharp quasiparticle peaks in the density of states. The system turns into a heavy fermion metal. But the 4$f$ electrons on the other Yb-ions (Yb2) remain localized and only weakly hybridized with conduction electrons. Superconductivity emerges in the heavy fermion phase when the Yb1-4$f$ Fermi surfaces develop a well nested quasi-two-dimensional structure. The nesting wave vector coincides with the antiferromagnetic wave vector of the localized spins, indicating possible involvement of heavy electrons and antiferromagnetic spin fluctuations. NaYbSe$_2$ might therefore be the 3rd Yb-based heavy fermion superconductor with a rather ``high" transition temperature ($T_c\sim 8\,$K) among all heavy fermion superconductors.\\

\section*{RESULTS AND DISCUSSION}
\noindent \textbf{Optimization of crystal structures}\\
\noindent The crystal structures of NaYbSe$_2$ at low and high pressures are compared in Fig.~\ref{fig1}a, both composed of edge-shared YbSe$_{6}$ and NaSe$_{6}$ octahedra. Yb ions form a layer structure of flat trianglular lattices with the shortest Yb-Yb distance in the $ab$ plane \cite{Jia2020}. The lattice parameters have been measured below 32 GPa in experiment \cite{Jia2020,Zhang2020}. For numerical calculations, we have used DFT+U to optimize the lattice parameters for both structures  \cite{Blaha2014,Perdew1996,Anisimov1997} and the results are compared with existing experimental data in Fig.~\ref{fig1}b. Their good agreement over the whole low-pressure range validates the starting point of our numerical calculations.\\

\begin{figure}[t]
\begin{center}
\includegraphics[width=0.48\textwidth]{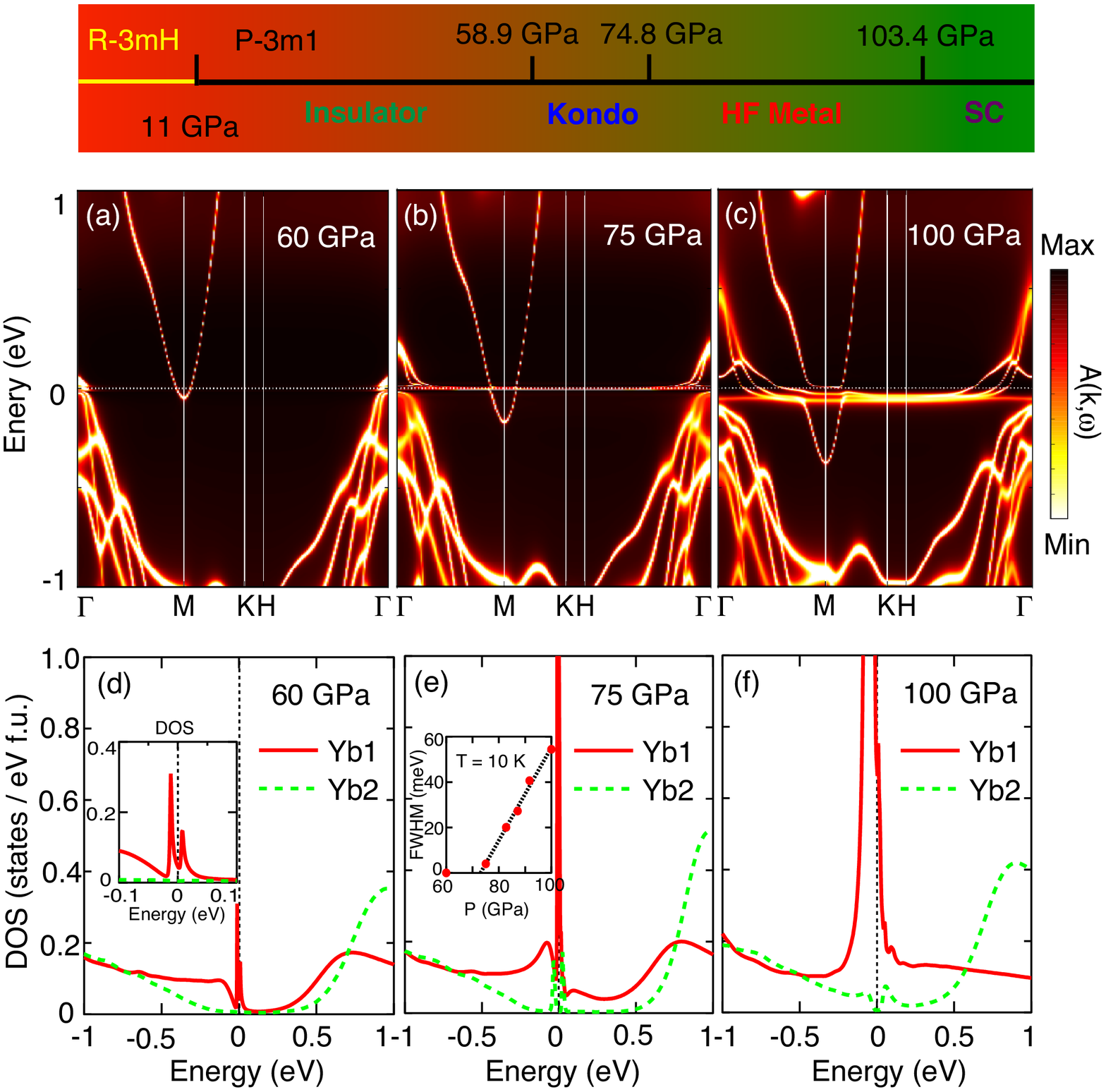}
\caption{{\bf Electronic phases of NaYbSe$_2$ under pressure.} The upper figure illustrates the predicted phases: the charge transfer or band insulator, the Kondo semimetal (Kondo), the heavy fermion metal (HF), and the superconductivity (SC). The pressures denote a rough boundary of these phases from experiment \cite{Jia2020}. {\bf a-c} The DFT+DMFT spectral functions at 10 K for 60, 75 and 100 GPa, respectively. {\bf d-f} The corresponding density of states on two inequivalent Yb ions, showing the development of a sharp quasiparticle peak of Yb1-$4f$ electrons. The inset in {\bf d} is an enlarged plot over a smaller energy window to show the pseudogap structure near the Fermi energy at 60 GPa. The inset in {\bf e} plots the full width at half maximum (FWHM) of the Yb1 quasiparticle peak as a function of pressure, where the solid line is a linear extrapolation giving a transition at about 73 GPa.}
\label{fig2}
\end{center}
\end{figure}

\noindent \textbf{Electronic structures at low pressures}\\
The electronic structures of NaYbSe$_2$ were obtained using DFT+DMFT with the hybridization expansion continuous-time quantum Monte Carlo (CT-HYB) as the impurity solver \cite{Werner2006,Haule2007}, taking into account both the spin-orbit coupling and electronic correlations. A Coulomb interaction $U=6$ eV was chosen to reproduce the correct charge gap of about 1.9 eV at ambient pressure estimated from absorption spectra measurements \cite{Liu2018}. Figure~\ref{fig1}c plots the calculated density of states and spectral function at $T=10$ K and ambient pressure. The Yb-4$f$ Hubbard bands are located at about -3.0 and 2.0 eV, respectively, with a clear charge gap between the Se-$3p$ bands and the Yb-$4f$ upper Hubbard bands. Hence NaYbSe$_2$ should be identified as a charge-transfer insulator at ambient pressure. Without DMFT, DFT+U calculations alone yield a much larger Yb occupation number and predict incorrectly a metallic ground state, in disagreement with experiment. The band structures at 20 GPa are compared in Fig.~\ref{fig1}d. We have now two inequivalent Yb-ions. Their $4f$ Hubbard bands become blurred, and the Yb-5$d$ bands move down closer to the Fermi energy to form a smaller normal band insulating gap with the slightly upshifted Se-$3p$ bands.\\

\noindent \textbf{The intermediate semi-metallic state}\\
Upon further increasing pressure, our calculations reveal a sequence of electronic phases, in good correspondence with the experimental observation. This is summarized in the upmost plot of Fig. \ref{fig2}. The first transition occurs near about 60 GPa, where the charge gap is closed and the spectral function reveals a semi-metallic state with a small electron pocket at the $M$ point from Yb-5$d$ bands and a hole pocket at $\Gamma$ from Se-$3p$ bands, as shown in Fig.~\ref{fig2}a. However, the Yb-4$f$ electrons are still localized at this pressure.  In Fig.~\ref{fig2}d, we see a tiny pseudogap structure near the Fermi energy in the $4f$ partial density of states, indicating a weak hybridization with conduction bands. Thus NaYbSe$_2$ should be better identified as a low carrier density Kondo system at this pressure. Consequently, there are no enough conduction electrons to screen the Yb-$4f$ local moments so that the resistivity should be dominated by incoherent Kondo scattering and exhibit typical logarithmic temperature dependence \cite{Luo2015,Chen2019}. This has indeed been observed in experiment between 58.9 and 74.8 GPa, but was initially attributed to weak localization \cite{Jia2020}. We argue that similar properties, including the logarithmic-in-$T$ resistivity and a pseudogap in the quasiparticle density of states, have previously been discussed in other low carrier density Kondo systems \cite{Luo2015,Chen2019,Nakamoto1995,Takabatake1998}. \\

\noindent \textbf{Heavy fermion metal at high pressure}\\
The second transition occurs close to 75 GPa beyond which flat heavy electron bands are seen to develop near the Fermi energy in Figs.~\ref{fig2}b and \ref{fig2}c. Correspondingly, a sharp quasiparticle peak appears at the Fermi energy in the Yb1-$4f$ density of states shown in Figs.~\ref{fig2}e and \ref{fig2}f due to the many-body Kondo effect. The system now turns into a good heavy fermion metal. To determine the exact transition point, we plot in the inset of Fig.~\ref{fig2}e the full width at half maximum (FWHM) of the Yb1-$4f$ quasiparticle peak. Its variation with pressure reflects tentatively the strength of the Kondo hybridization \cite{Krawiec2006}. A linear extrapolation to zero suggests that the transition occurs at about 73 GPa, which separates the low carrier density Kondo phase and the heavy fermion metal. Thus, it marks a delocalization transition of the Yb1-$4f$ electrons. Interestingly, experiment did observe a non-Fermi liquid state with linear-in-$T$ resistivity near 74.8 GPa and beyond that a gradual crossover to the Fermi liquid at 126 GPa \cite{Jia2020}. This agrees with our theory and strongly suggests the possible existence of a delocalization quantum critical point at zero temperature. However, the broad crossover region is quite unusual and might be associated with the special layer structure of NaYbSe$_2$ to be discussed later. Nevertheless, the insulator-to-metal transition is not a simple Mott transition, but a two-stage process first from a normal band insulator to a semimetal with incoherent Kondo scattering and then to a heavy fermion metal. Similar transition and non-Fermi liquid property have also  been reported in  the low carrier density Kondo system CeNi$_{2-\delta}$(As$_{1-x}$P$_{x}$)$_{2}$ \cite{Chen2019}.

It may be interesting to mention that pressure seems to favor the heavy fermion phase in NaYbSe$_2$, while in many other Yb-based heavy fermion compounds, the Kondo effect is typically weakened under pressure and gives in to long-range magnetic orders because of the upshift of the Yb-$4f$ hole (upper-Hubbard) bands further away from the Fermi energy,  making Yb-$4f$ electrons more localized. In NaYbSe$_2$, our results show that the localized Yb-$4f$ levels are already away from the Fermi energy and the major effect of pressure is to increase the carrier density and enhance the hybridization as will be discussed below. Such anomalous increase of the Kondo effect with pressure has previously been observed in Yb$_2$Ni$_2$Al and YbCu$_{2}$Si$_{2}$ \cite{Winkelmann1998,Winkelmann1999} but attributed to the electron-lattice coupling \cite{Goltsev2005}. Here NaYbSe$_2$ provides an interesting alternative for further investigation. \\

\begin{figure}
\begin{center}
\includegraphics[width=0.49\textwidth]{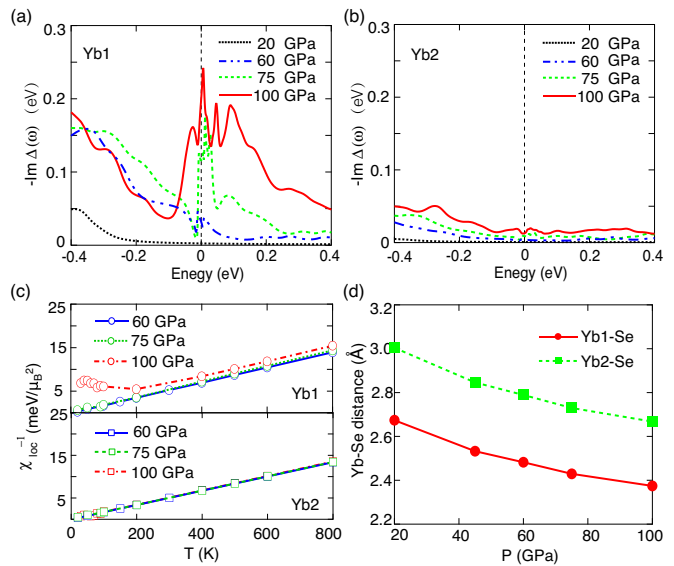}
\caption{{\bf Different properties of the two inquivalent Yb-ions.} {\bf a, b} Pressure evolution of the minus imaginary part of the hybridization function in real frequency on Yb1 and Yb2 ions from 20 to 100 GPa, obtained by DFT+DMFT calculations at 10 K. {\bf c} Comparison of static local spin susceptibility of Yb1 and Yb2 $4f$-electrons at 60, 75, and 100 GPa. {\bf d} Comparison of the Yb-Se distances as a function of pressure from 20 to 100 GPa.}
\label{fig3}
\end{center}
\end{figure}

\noindent \textbf{Alternating heavy fermion and QSL layers}\\
To have a better understanding of the peculiarity of NaYbSe$_2$, we compare in Figs.~\ref{fig3}a and \ref{fig3}b the imaginary part of the Yb-$4f$ hybridization function at four typical pressures. Quite remarkably, this reveals an unexpected but substantial difference between the two inequivalent Yb ions. While no hybridization is observed near the Fermi energy on both ions at low pressure (20 GPa), it begins to develop on Yb1 at 60 GPa and becomes large above 75 GPa, but on Yb2 the hybridization remains weak over the whole pressure range and shows no significant enhancement. These suggest that only Yb1 ions are responsible for the flat bands and heavy fermion properties at high pressure, while Yb2-$4f$ electrons remain localized and weakly interacting with conduction electrons. These observations are further confirmed in the local spin susceptibility, $\chi_{zz}(\omega)= \left< S_{z}(\tau)S_{z}(0)\right>_\omega$, calculated using the CTQMC impurity solver. Here $S_z$ is the local spin operator of the Yb-$4f$ electrons and $\tau$ is the imaginary time. As shown in Fig.~\ref{fig3}c, the static susceptibility of Yb1 deviates from the Curie-Weiss law at low temperature above 75 GPa due to the Kondo hybridization, but that of Yb2 always follows the Curie-Weiss law and keeps the local moment behavior. This explains the absence of quasiparticle resonance in the Yb2-$4f$ density of states plotted earlier in Fig.~\ref{fig2}f. To understand these huge differences between Yb1 and Yb2 ions, we compare in Fig.~\ref{fig3}d their local environment, namely their respective Yb-Se bond distances. With increasing pressure from 20 to 100 GPa, the Yb1-Se distance is seen to decrease from 2.67 to 2.37 $\AA{}$,  while the Yb2-Se distance decreases from 3.0 to 2.66 $\AA{}$. Hence the Yb2-Se distance at 100 GPa is roughly equal to the Yb1-Se distance at 20 GPa, explaining why the Yb2-$4f$ electrons are still localized at such high pressure. We thus conclude that the high-pressure structure of NaYbSe$_2$ contains alternating Yb1 heavy fermion layers and Yb2 QSL layers, a very peculiar structure for studying the interplay between the two exotic quantum many-body states. It is wondering if this might be responsible for the broad crossover to the Fermi liquid behavior observed in the resistivity \cite{Jia2020}. \\

\begin{figure}
\begin{center}
\includegraphics[width=0.48\textwidth]{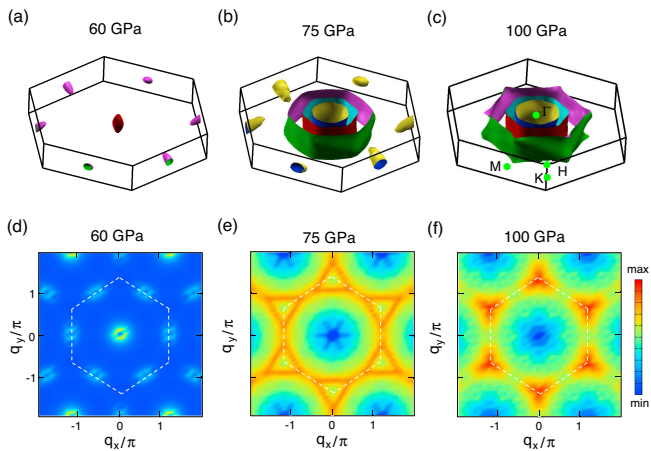}
\caption{{\bf Fermi surface topology and nesting with pressure.} {\bf a-c} Comparison of the DFT+DMFT calculated Fermi surfaces at 60, 75 and 100 GPa obtained at 10 K. {\bf d-f} The corresponding Lindhard susceptibility (real part), showing a good nesting property only beyond 100 GPa.}
\label{fig4}
\end{center}
\end{figure}

\begin{table}[t]
\caption{\label{tab1} {\bf Spin singlet pairing states from irreducible representations of the point group $D_{3d}$ of Yb1 ions.} The line$^*$ means possible accidental line nodes for certain parameters of the basis functions and otherwise fully gapped.\\}
\centering
\begin{tabular}{|c|c|c|}
\hline
IR & singlet gap functions $\Delta({\bf k})$ & nodes\\
\hline
$A_{1g}$ & $1,\ k_x^2+k_y^2,\ k_z^2$ & line$^*$ \\
$A_{2g}$ & $k_xk_z(k_x^2-3k_y^2),\ k_yk_z(k_y^2-3k_x^2)$ & line \\
$E_{g}$ & $(k_x^2-k_y^2,k_xk_y),\ (k_xk_z,k_yk_z)$ & line$^*$ \\
\hline
\end{tabular}
\end{table}

\noindent \textbf{Superconductivity}\\
Experimentally, superconductivity in NaYbSe$_2$ emerges above 103 GPa in the heavy Ferm liquid phase \cite{Jia2020}. To gain some insight on its pairing mechanism, we compare in Figs.~\ref{fig4}a-c the Fermi surface structures at 60, 75 and 100 GPa, respectively. At 60 GPa, the Fermi surfaces only consist of small electron and hole pockets centered at M and $\Gamma$ points. The small Fermi volume reflects a low carrier density of the semi-metallic phase for incoherent Kondo scattering. At 75 GPa, the Fermi surfaces become larger with three cylinders around $\Gamma$ and a small electron pocket around the M point. At 100 GPa, the cylindrical Fermi surfaces continue to grow but the electron pocket at M is gapped out and diminishes due to strong hybridization. We find superconductivity emerges as the Fermi surfaces become well nested and quasi-two-dimensional. This is seen in Figs.~\ref{fig4}d-f, where we plot the real part of the Lindhard susceptibility derived from DFT+DMFT Fermi surfaces. Only beyond 100 GPa, we find a well-defined nesting wave vector. It is thus speculated that the Fermi surface nesting and its resulting spin density wave fluctuations may play a major role in the Cooper pairing. Moreover, the nesting pattern is in close resemblance to that of the spin excitation spectra of the Heisenberg model on the same lattice as revealed by neutron scattering experiment at ambient pressure \cite{Dai2021}. Considering that the Yb2 layers remain in a QSL state, it may be natural to wonder if their associated magnetic fluctuations may also be involved in the electron pairing. At this stage, phonons also cannot be excluded. But in any case, since NaYbSe$_2$ is in a heavy fermion state at this pressure, the itinerant heavy Yb1-4$f$ quasiparticles must participate in the superconductivity, making it a candidate heavy fermion superconductor. If this is the case, it would be the 3rd Yb-based heavy fermion superconductors, in addition to YbRh$_2$Si$_2$ of $T_c\sim2$ mK \cite{Schuberth2016} and $\beta$-YbAlB$_4$ of $T_c\sim80$ mK \cite{Nakatsuji2008}. The transition temperature of about 8 K in NaYbSe$_2$ is strikingly high among all heavy fermion superconductors except Pu-115, probably suggesting a potential cooperation of the Fermi surface nesting, spin fluctuations and phonons. For such a Fermi surface topology with the antiferromagnetic nesting wave vector, group-theoretical analyses favor a spin singlet pairing state with gap symmetries listed in Table~\ref{tab1}. We have either a nodeless  $s$-wave pairing ($A_{1g}$) or a two-component $d$-wave pairing ($E_g$), with possible accidental line nodes for appropriate combination of the basis functions, or a less probable $g$-wave gap ($A_{2g}$) with loop nodes in the $k_z=0$ plane. More elaborate experiments are needed for better clarification but would be extremely difficult under such high pressures. It will be important if superconductivity can be found at lower pressures in other members of the same family through chemical substitution. \\

\noindent \textbf{Summary}\\
Our calculations reveal a sequence of electronic phases in pressurized QSL candidate NaYbSe$_2$. We find that it is a charge-transfer insulator at ambient pressure and turns into a low carrier density Kondo system with semi-metallic band structures close to 60 GPa and a good heavy fermion metal above 73 GPa. In between, we predict a delocalization quantum critical point of the Yb1-$4f$ electrons that may explain the observed non-Fermi liquid with linear-in-$T$ resistivity, while the Yb2-$4f$ electrons remain localized over the whole pressure range. The lattice therefore has a special structure constructed with alternating Yb1 heavy fermion layers and Yb2 QSL layers. Superconductivity emerges in the heavy fermion metal with well nested quasi-two-dimensional Fermi surfaces, making NaYbSe$_2$ possibly the 3rd Yb-based heavy-fermion superconductor with a record-high $T_c$. This suggests the possible involvement of both spin fluctuations and phonons as the pairing glues. Our work will be a useful guide for future experimental study of NaYbSe$_2$ and its other family members.\\

\section*{METHODS}
DFT calculations were performed using the full-potential augmented plane-wave plus local orbital method as implemented in the WIEN2k package \cite{Blaha2014} and the Perdew-Burke-Ernzerhof exchange-correlation potential \cite{Perdew1996}. The lattice parameters were optimized using DFT+U with an effective Coulomb interaction of 6 eV \cite{Anisimov1997}. The plane-wave cutoff $K_{\rm max}$ was chosen to make $R_{\rm MT} \times K_{\rm max}=8.0$. All calculations were performed on a grid of 1000 k-points in the Brillouin zone. The spin-orbit coupling was included explicitly. 

DFT+DMFT calculations were employed to treat the electronic correlations of Yb-$4f$ orbitals. The hybridization expansion continuous-time quantum Monte Carlo (CT-HYB) was used as the impurity solver \cite{Werner2006,Haule2007}. Yb-$4f$ local orbitals were constructed using projectors with an energy window from -10 to 10 eV relative to the Fermi energy. The Coulomb interaction was chosen to be $U=6$ eV and $J_{H}=0.7$ eV. A nominal double-counting scheme with $n_f=13.0$ was used. The self-energy in real frequency was obtained by analytic continuation based on the maximum entropy \cite{Haule2010,Jarrell1996}.\\

\section*{ACKNOWLEDGEMENTS}
This work was supported by the National Natural Science Foundation of China (NSFC Grant No. 11774401,  No. 11974397), the National Key R\&D Program of MOST of China (Grant No. 2017YFA0303103), the Strategic Priority Research Program of the Chinese Academy of Sciences (Grant No. XDB33010100), and the Youth Innovation Promotion Association of CAS.

\end{document}